\documentclass[12pt]{elsart}

\usepackage[latin1]{inputenc}
\usepackage{amsmath}
\usepackage{amsfonts}
\usepackage{amssymb}
\usepackage{ulem}
\usepackage[dvips]{graphics}
\usepackage{epsfig}

\begin{document}

\begin{frontmatter}

\title{A Model of Coupled-Maps for Economic Dynamics}
\author[jrs]{J.R. S\'{a}nchez},
\ead{jsanchez@fi.mdp.edu.ar}
\author[rlr]{R. L\'{o}pez-Ruiz}
\ead{rilopez@unizar.es}

\address[jrs]{Facultad de Ingenier\'{\i}a, Universidad Nacional de Mar del Plata, \\
Avda. J.B. Justo 4302, 7600 - Mar del Plata, Argentina,}
\address[rlr]{Facultad de Ciencias, DIIS and BIFI,
Universidad de Zaragoza, \\ 50009 - Zaragoza, España.}

\begin{abstract}
An array system of coupled maps is proposed as a model for 
economy evolution. The local dynamics of each map or agent is controlled by two parameters.
One of them represents the growth capacity of the agent
and the other one is a control term representing the local environmental pressure 
which avoids an exponential growth. The asymptotic state of the system evolution 
displays a complex behavior. The distribution of the maps values in this final regime 
is of power law type. In the model, inequality emerges 
as a result of the dynamical processes taking place in the microscopic scales.

\end{abstract}


\end{frontmatter}

Inequality in the richness distribution is a fact in each
economic activity. The origin of such behavior seems to be caused by
the interaction of the macro with the microeconomy. Here we propose
a simple spatio-temporal model for economy evolution where inequality emerges 
as a result of the dynamical processes taking only place on the microscopic scale.
That is, the microeconomy fully determines the macroeconomic characteristics of the system.

The model is composed by $N$ interacting agents representing a company, country or other economic entity.
Each agent, identified by an index $i=1 \cdots N$, is characterized by a real, scalar degree of freedom,
$x_i \in [0,\infty]$ denoting the {\it strength, wealth or richness} of the agent.
The system evolves in time $t$ synchronously. Each agent updates its $x_i^t$ value according 
to its present state and the value of its nearest-neighbors. Thus, the value of $x_i^{t+1}$ 
is given by the product of two terms; the {\it natural growth} of the agent $r_ix_i^{t}$
with positive local ratio $r_i$, and a {\it control term} that limits this growth with respect to
the local field $\Psi_i^t=\frac{1}{2}(x_{i-1}^{t}+x_{i+1}^{t})$ through a negative 
exponential with parameter $a_i$:   
\begin{equation}
x_i^{t+1} = r_i\:x_i^{t}\: \exp(-\mid x_i^{t}-a_i\Psi_i^t\mid).
\end{equation}
The parameter $r_i$ represents the {\it capacity} of the agent to become richer
and the parameter $a_i$ describes the local {\it selection pressure} \cite{ausloos}.
This means that the largest possibility of growth for the agent is obtained
when $x_i\simeq a_i\Psi_i^t$, i.e., when the agent has reached some kind of adaptation 
to the local environment. In this note, for the sake of simplicity, we concentrate our 
interest in a homogeneous system with a constant capacity $r$ and a constant
selection pressure $a$ for the whole array of sites. 
 
If all the agents start with the same wealth, the index $i$ can be removed,
$x_i^t=x^t$ and $\Psi_i^t=x^t$,
and the global evolution reduces to the following map,
\begin{equation}
x^{t+1}=r x^{t} \exp(-\mid (1-a)x^{t}\mid).
\end{equation}
The above map can be easily analyzed by standard techniques.
For $r<1$ the system relaxes to zero and for $r>1$ the dynamics
can be self-sustained deriving toward different types of attractors.
It displays all kind of bifurcations known
for this type of maps \cite{schuster}, except, evidently, 
for the singular case $a=1$. For instance, when $r>1$ 
the fixed point is $x_0=\log r/\mid 1-a\mid$.
This point becomes unstable by a flip bifurcation for $r=e^2$.
For increasing $r$ the whole period doubling cascade and other
complex dynamical behavior are obtained.
However, it can be shown that such evolving uniform states are unstable.
When a perturbation is introduced in the initial uniform state or, in general,
when the initial condition is a completely random one,
the asymptotic dynamical state of the system is found to be
more complex.

In Fig. 1 the after-transient spatial mean value of the field $x_i$ is shown as a function of the
capacity parameter $r$ using a selection pressure $a=0.8$. It can be seen that
for values of $r \simeq 1$ the mean reaches a uniform, constant limiting value, but for greater
values of $r$ the dynamics becomes spatio-temporally complex.

In Fig. 2, the frequency of agent's richness in the asymptotic state of the system, is log-log plotted. 
As it can be seen , it scales as a Pareto like power law ~\cite{lorenz,reed}.
This scaling law, expressed as $P(s) \sim s^{\:\beta}$ with $\beta = -2.21$, is of the same type to those ones
directly obtained from actual economy data ~\cite{stanley}.

\newpage
\begin{figure}[ht] 
\begin{center}
{\epsfig{file=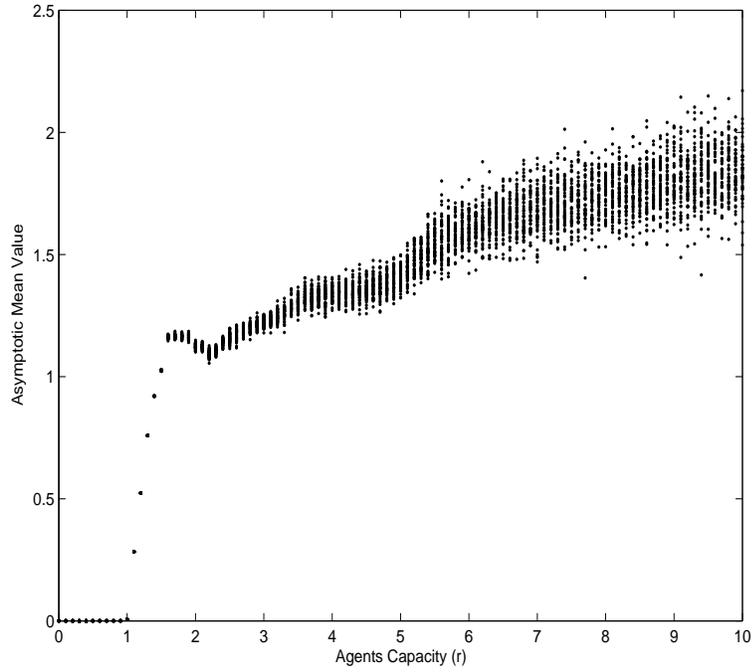,height=9cm,width=10cm}}
\end{center}
\caption{Asymptotic spatial mean value of the field $x_i$ as a function of the
capacity parameter $r$, using a selection pressure $a=0.8$.}
\end{figure}
\begin{figure}[ht] 
\begin{center}
{\epsfig{file=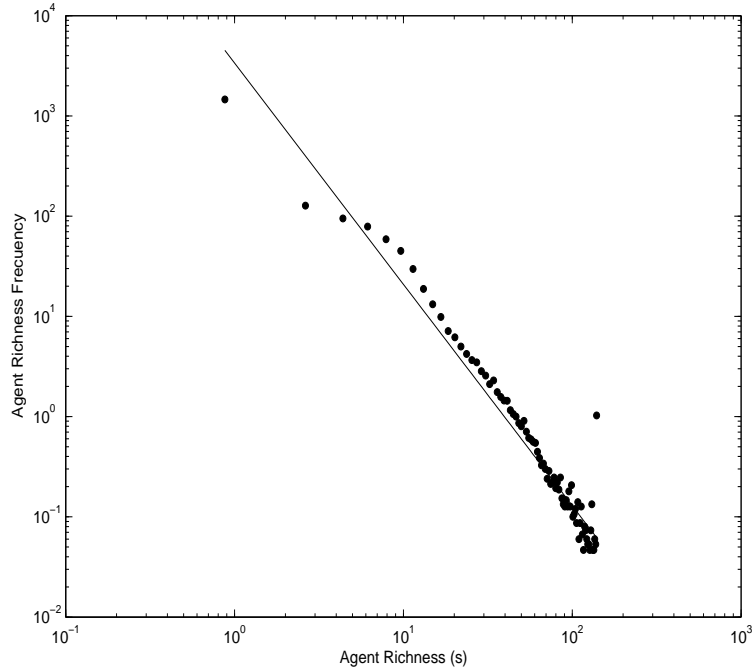,height=9cm,width=10cm}}
\end{center}
\caption{Log-log plot of the frequency of agent's richness in the asymptotic state of the system.}
\end{figure}


\newpage
\begin{thebibliography}{99}

\bibitem{ausloos} M. Ausloos, P. Clippe and A. Pekalski, 
``Simple Model for the Dynamics of Correlations in the Evolution of 
Economic Entities under Varying Economic Conditions",
Physica A {\bf 324}, 330-337 (2003) .

\bibitem{schuster} H.G. Schuster, {\it Deterministic Chaos}, Physik-Verlag, Weinheim (1984). 

\bibitem{lorenz} M.O. Lorenz,``Methods of Measuring the Concentration of Wealth", 
Publications of the American Statistical Association {\bf 9}, 209-219 (1905).

\bibitem{reed} W. J. Reed,``The Pareto, Zipf and other Power Laws",  
Econ. Lett. {\bf 74}, 15 -19 (2001); 
$http://linkage.rockefeller.edu/wli/zipf/reed01_el.pdf$.

\bibitem{stanley} L.A. Nunes-Amaral et al., 
``Scaling Behavior in Economics: I. Empirical Results for Company Growth",
J. Phys. I France {\bf 7}, 621-623 (1997); 
and references therein. 

\end{thebibliography}
\end{document}